\documentstyle[prc,preprint,tighten,,aps,epsf]{revtex}
\begin{document}
\draft
\preprint{\vbox{\noindent 
 \hfill LA-UR-98-173\\
 \null\hfill nucl-th/9802004}}
\title{Large-space cluster model calculations for 
the $\bbox{^3}$He($\bbox{^3}$He,2p)$\bbox{^4}$He
and $\bbox{^3}$H($\bbox{^3}$H,2n)$\bbox{^4}$He reactions}
\author{Attila Cs\'ot\'o$^{a,b,c}$, Karlheinz Langanke$^b$}
\address{$^a$Theoretical Division, Los Alamos National 
Laboratory, Los Alamos, NM 87545, USA \\
$^b$Institute for Physics and Astronomy, Aarhus University,
DK-8000 Aarhus, Denmark \\
$^c$Department of Atomic Physics, E\"otv\"os University, 
P\'azm\'any P\'eter s\'et\'any 1/A, H--1117 Budapest, Hungary}
\date{\today}

\maketitle

\begin{abstract}
The $^3{\rm He}({^3{\rm He}},2p){^4{\rm He}}$ and 
$^3{\rm H}({^3{\rm H}},2n){^4{\rm He}}$ reactions are
studied in a microscopic cluster model. We search for
resonances in the $^3$He+$^3$He and $^4{\rm He}+p+p$ 
channels using methods that treat the two- and 
three-body resonance asymptotics correctly. Our results 
show that the existence of a low-energy resonance or virtual 
state, which could influence the $^7$Be and $^8$B solar 
neutrino fluxes, is rather unlikely. Our calculated $^3{\rm 
He}({^3{\rm He}},2p){^4{\rm He}}$ and $^3{\rm H}({^3{\rm 
H}},2n){^4{\rm He}}$ cross sections are in a good general 
agreement with the experimental data.
\end{abstract}


\narrowtext 

\section{Introduction}

One way to test models
of the solar interior is by observing the 
neutrinos generated by the nuclear reaction 
network which is the solar energy source \cite{SNexp}. As a 
striking and exciting fact, all terrestrial solar neutrino
experiments observe fewer neutrinos than predicted by
standard solar models \cite{SNtheo}. The picture which 
emerges from the various experiments is \cite{Hata} that the 
$^8$B flux is about half of its predicted value 
($\phi_8=0.4\phi_8^{SSM}$), while the $^7$Be neutrinos 
appear to be completely missing ($\phi_7=0$). Here SSM 
refers to the standard solar model of Bahcall 
\cite{SNtheo}. The $^7$Be nuclei, which are the seeds of 
both the $^7$Be and $^8$B neutrinos, are produced in the 
$^4{\rm He}({^3{\rm He}},\gamma){^7{\rm Be}}$ reaction. This 
reaction competes with the $^3{\rm He}({^3{\rm He}},2p){^4{
\rm He}}$ process, which is the final  step of the first 
branch of solar hydrogen burning (ppI chain) \cite{Bahcall}. 
If the cross section of the $^3{\rm He}({^3{\rm He}},2p){^4{
\rm He}}$ reaction were larger than believed then the 
probability of the $^4{\rm He}({^3{\rm He}},\gamma){^7{\rm 
Be}}$ branch would be smaller, and hence the $\phi_8$ and 
$\phi_7$ fluxes would be suppressed (without, however,
significantly affecting the unexpected $\phi_7 / \phi_8$ 
ratio deduced from the neutrino experiments 
\cite{Raghavan}). To increase the $^3{\rm He}({^3{\rm 
He}},2p){^4{\rm He}}$ reaction rate, a hypothetical 
resonance at low energies had been suggested \cite{Fowler} 
and looked for in various experiments (see \cite{LUNA} for 
the most recent work).

The importance of the $^3{\rm He}({^3{\rm He}},2p){^4{\rm 
He}}$ reaction has led to continued experimental efforts to
measure the cross section down to solar energies. In the
latest experiment \cite{LUNA} the LUNA collaboration 
measured the cross section down to $E_{cm}=20.76$ keV, which 
is well within the region of the most effective solar 
energies (Gamow window). Although, they do not see any 
evidence of a possible resonance, the existence of such a 
state at still lower energies cannot be {\it a priori} ruled 
out yet. In the present work we study the $^3{\rm He}({^3{
\rm He}},2p){^4{\rm He}}$ and the mirror $^3{\rm H}({^3{\rm 
H}},2n){^4{\rm He}}$ reactions in a microscopic cluster 
model. We search for signs of possible resonances and study 
the energy dependence of the reaction cross sections.

\section{Model}

The $^3{\rm He}({^3{\rm He}},2p){^4{\rm He}}$ and $^3{\rm 
H}({^3{\rm H}},2n){^4{\rm He}}$ reactions have already been studied
previously within  microscopic cluster models \cite{Typel,Desc}. We
use the same model, but with extended and hence more realistic model spaces. 
Additionally we put special emphasis on
the description of the few-body dynamics. Our model starts
out with a resonating group model (RGM) wave function for 
the $\{{^4{\rm He}}+p+p,{^3{\rm He}}+{^3{\rm He}}\}$
coupled system.  
\begin{eqnarray}
\Psi^{J\pi}&=&\sum_{S,l_1,l_2,L}
{\cal A}\Bigg \{\bigg [ \Big [\Phi^\alpha(\Phi^p\Phi^p)
\Big ]_S
\chi_{[l_1l_2]L}(\mbox{\boldmath
$\rho $}_1,\mbox{\boldmath $\rho $}_2)
\bigg ]_{JM} \Bigg \}\cr 
&+&\sum_{S,l_1,l_2,L}{\cal A}\Bigg
\{\bigg [ \Big [\Phi^p(\Phi ^\alpha\Phi^p) \Big ]_S
\chi_{[l_1l_2]L}(\mbox{\boldmath
$\rho $}_1,\mbox{\boldmath $\rho $}_2)
\bigg ]_{JM} \Bigg \} 
+{\cal A}\bigg \{\Big [\Phi^h\Phi^h
\chi_{l_3}(\mbox{\boldmath
$\rho $})\Big ]_{JM} \bigg \}.
\label{wf}
\end{eqnarray}
Here ${\cal A}$ is the intercluster antisymmetrizer, 
the $\Phi$ cluster internal states are translationally 
invariant harmonic oscillator shell model states ($\alpha 
=$~$^4$He and $h =$~$^3$He), the \mbox{\boldmath $\rho $} 
vectors are the different intercluster Jacobi coordinates, 
$l_1$ and $l_2$ are the angular momenta of the two relative 
motions, $L$ and $S$ are the total orbital angular momentum 
and spin, respectively, $J$ is the total angular momentum, 
$\pi=(-1)^{l_1+l_2}=(-1)^{l_3}$ is the parity, and [...] 
denotes angular momentum coupling. The sum over $S,l_1,l_2$, 
and $L$ includes all physically relevant angular momentum 
configurations. Using (\ref{wf}) as a trial function in the 
six-nucleon Schr\"odinger equation, we arrive at an equation 
for the intercluster relative motion functions $\chi$. For 
the mirror $\{{^4{\rm He}}+n+n,{^3{\rm H}}+{^3{\rm H}}\}$
system we use a wave function similar to (\ref{wf}). The 
harmonic oscillator size parameters of the internal cluster  
states are chosen to stabilize the cluster energies. We use
the Minnesota (MN) effective interaction \cite{MN} in all 
calculations. It puts the threshold of the $^3{\rm
He}+{^3{\rm He}}$ and $^3{\rm H}+{^3{\rm H}}$ thresholds at
17.11 MeV and 15.57 MeV energies, respectively relative to 
the $^4{\rm He}+N+N$ threshold. Like in the previous studies
\cite{Typel,Desc}, these values are 
significantly larger than the experimental thresholds (12.86 MeV and 
11.33 MeV, respectively). One might expect that the incorrect
reproduction of the threshold energies will effect the relative weight
of the $^4{\rm He}+N+N$ channels in the full wave function. We will
briefly discuss this problem below.

The relative motions are the most important degrees of
freedom in the problem, so special care has to be taken in
order to ensure that their dynamics are described properly.
What makes the description of the $^3{\rm He}({^3{\rm 
He}},2p){^4{\rm He}}$ and its mirror reactions difficult is the
fact that there are three particles in the final state
continuum. Currently we cannot treat the full three-body
continuum problem properly, thus our model is only an
approximate description of the reactions. However, 
the
existence of a low-energy resonance in the $^3{\rm He}({^3{\rm 
He}},2p){^4{\rm He}}$ reaction is a question which can be studied in
a rigorous way. If such a resonance existed, it has to
originate from either the  $^3{\rm He}+{^3{\rm
He}}$ or  the $^4{\rm He}+p+p$ channels. As we will show below, one can
investigate the existence of such a resonance in both channels
separately and properly.
Although we can treat the scattering continuum only approximately, we
will nevertheless calculate the low-energy
reaction cross sections of the $^3{\rm He}({^3{\rm 
He}},2p){^4{\rm He}}$ and its mirror reaction.

\section{Searching for resonances in 
$\bbox{^4}$H\lowercase{e+p+p} and 
$\bbox{^3}$H\lowercase{e}+$\bbox{^3}$H\lowercase{e}}

In order to avoid any ambiguity in the recognition of a
resonance in real-energy observables, we search for
resonances on the complex energy plane. Resonances are
defined as the complex-energy solutions of the
Schr\"odinger equation, which correspond to the
singularities of the scattering matrix. Thus, we search for
poles of both the $^3{\rm He}+{^3{\rm He}}$ and $^4{\rm He}
+p+p$ $S$ matrices.

Interestingly, the $^4{\rm He}+p+p$ case is easier to deal
with, despite its three-body nature. The reason is that a
low-energy resonance in the $^3{\rm He}+{^3{\rm He}}$
channel corresponds to a high-lying narrow state in 
the $^4{\rm He}+p+p$ channel, which, if it exists, can be easily
identified.
In order
to be able to describe three-body resonances, we apply the
complex scaling method (CSM). This method has already  
been used previously to search for resonances in $^6$He, 
$^6$Li, and $^6$Be in an $\alpha+N+N$ model \cite{soft}. 
However, the search had been restricted to low-lying 
resonances below the $^3$He+$^3$He threshold and therefore 
does not shed light on the problem at hand here. In the 
present search we concentrate on the high-energy region 
around the $^3{\rm He}+{^3{\rm He}}$ channel threshold. In 
the CSM the complex scaling transformation is performed on 
the original Hamiltonian. This transformation acts in
coordinate space on a function $f({\bf r})$ as 
\begin{equation}
\widehat{U}(\theta)f({\bf r})=e^{3 i \theta /2}f({\bf 
r}e^{i\theta}),
\label{cs}
\end{equation}
where $\theta$ is the parameter of the transformation. For 
real $\theta$ values the $\widehat{U}(\theta)$
transformation results in a rotation into the complex 
coordinate plane. The spectrum of the complex-scaled
Hamiltonian 
\begin{equation}
\widehat{H}_\theta=\widehat{U}(\theta)\widehat{H}
\widehat{U}^{-1}(\theta)
\end{equation}
is connected to the spectrum of the original $\widehat H$
by the following theorem \cite{ABC}: (i) the bound 
eigenstates of $\widehat{H}$ are eigenstates of 
$\widehat{H}_\theta$, for any value of $\theta$ within 
$0\leq \theta \leq \pi/2$; (ii) the continuous spectrum of 
$\widehat{H}$ is rotated by an angle $2\theta$; (iii) the 
complex generalized eigenvalues of $\widehat{H}_\theta$,
$\varepsilon_{\text{res}}=E_{\rm r}-i\Gamma/2$
(with $E_{\rm r},\, \Gamma >0$), belong to its
proper spectrum, with square-integrable eigenfunctions,
provided $2\theta>|{\rm arg}\, \varepsilon_{\text{res}}|$.
These complex eigenvalues coincide with the $S$-matrix
pole positions. 

We would like to emphasize that this method treats
three-body resonances in a rigorous way. The only
approximation we make in the present work is that we solve
the complex-scaled Schr\"odinger equation on a finite
basis. Namely, we assume that $\chi$ in (\ref{wf}) can be
expanded in terms of products of Gaussian functions, 
like $\rho_1^{l_1} \exp [-(\rho_1/\gamma_i)^2]Y_{l_1m_1}(
\widehat\rho_1)\cdot \rho_2^{l_2}\exp [-(\rho_2/
\gamma_j)^2]Y_{l_2m_2}(\widehat \rho_2)$, where $l_1$ and 
$l_2$ are the angular momenta in the two relative motions, 
respectively, and the widths $\gamma$ of the Gaussians are 
the parameters of the expansion. The expansion coefficients
are determined from the projection equation $\langle\delta \Psi_\theta|
\widehat{H}_\theta-\varepsilon| \Psi_\theta\rangle=0$.
As the complex-scaled wave function of
a resonance is square-integrable, our finite basis
approximation works  for  resonances as well as for  bound
states. 

As an illustrative example of our results, Fig.\ 1 shows
the spectrum of the complex-scaled Hamiltonian for the
$J^\pi=0^+$ state of $^4{\rm He}+p+p$. We choose a small 
rotation angle $\theta$ in order to avoid any numerical 
instability. One can see that the well-known ground state 
resonance of $^6$Be is reproduced. A low-energy resonance 
in the $^3{\rm He}({^3{\rm He}},2p){^4{\rm He}}$ system with small 
width (which would be relevant for the astrophysical 
problem), if existed, would be situated above ${\rm
Re}(E)\approx10$ MeV, close to the real energy axis. 
There is no such narrow state present in our model. We do 
not find such a state in other $J^\pi$ partial waves 
either. Our results agree with other calculations performed 
in a three-body $^4{\rm He}+N+N$ model assuming 
structureless $^4$He \cite{Aoyama}. As Ref.\ 
\cite{Aoyama,Janecke} shows, all high-lying resonances in 
the $^4{\rm He}+N+N$ systems have large widths. 

In the $^3{\rm He}+{^3{\rm He}}$ channel we cannot use the
complex scaling method. A very low-energy resonance would
always be mixed up with the rotated continuum, making its
unambiguous identification hopeless. Here we use a direct
analytic continuation of the $S$ matrix to complex energies
\cite{ac}. We can use this method because, unlike in the
case of the three-body $^4{\rm He}+p+p$ system, the two-body
scattering wave functions can easily be generated with the
correct asymptotics. To calculate these scattering wave
functions we use the Kohn-Hulth\'en variational method of
Ref.\ \cite{Kamimura}. Once again, our method treats resonances 
in a rigorous way. 

We have searched for low-energy narrow resonances in 
$^3{\rm He}+{^3{\rm He}}$ and found none. We must note,
however, that our $^3{\rm He}+{^3{\rm He}}$ model may be
too simple. In the $J^\pi=0^+$ state, which is the most
likely candidate to have a narrow resonance, the singlet
$S$-wave $N-N$ interaction has a dominant effect because
the Pauli principle forces the unpaired neutrons inside the
$^3$He clusters to form a ${^1S}_0$ state between them. In
the $^3{\rm He}+{^2{\rm H}}$ system the role of the triplet
forces is known to be very strong, causing the existence
of the low-lying $3/2^+$ resonance \cite{Bluge}. The
triplet forces, which have negligible effects in our
present model, could play a role if the small $D$-state admixture
of $^3$He were to be considered. In such a case one can have a contribution
from coupling between the
$\bigg \{\Big
[(1,1/2)1/2,0\Big ]1/2;\Big [(1,1/2)3/2,2\Big ]1/2\bigg
\}$ configurations in the $^3$He+$^3$He system. Here
the $\Big [(S_d,S_p)S,l\Big ]I$ coupling scheme is used,
with $S_d$ and $S_p$ being the deuteron and proton spins,
respectively, $S$ is the total intrinsic spin, $l$ is the
angular momentum between the deuteron and the proton, $I$ 
is the total spin of $^3$He, and the brackets denote
angular momentum coupling. Such a model would require a
four-cluster description of $^3{\rm He}+{^3{\rm
He}}=(d+p)+(d+p)$, which is beyond the scope of the present
work. The effects of the internal $D$ states in $^3$He on
the $^3{\rm He}+{^3{\rm He}}$ system will be studied in the
future \cite{Varga}.

In many respects $^3{\rm He}+{^3{\rm He}}$ is similar to
the $n+n$ system. We know that there is a virtual
(antibound) state present in the $n+n$ system, with the wave number 
$k=-i\gamma\ \
(\gamma>0)$ and energy $E=-E_V\ \ (E_V>0)$ 
of the $S$-matrix pole, respectively. Such a state results
in a cross section which is divergent at the unphysical
negative pole energy and behaves as $\sigma\sim1/(E+E_V)$
at positive energies. The effect of such a hypothetical
state in $^3{\rm He}+{^3{\rm He}}$ on the $^3{\rm He}({^3{
\rm He}},2p){^4{\rm He}}$ cross section was discussed in
Ref.\ \cite{innocom}. It would lead to a cross section that
increases with decreasing energy, mimicking the effect of
electron screening. A closer look at the problem shows that
pure virtual states (with pure imaginary wave number)
cannot be present in Coulombic systems \cite{H3}. The
Coulomb interaction creates two poles from the one
virtual-state pole and moves them away from the imaginary
$k$-axis to $k=\pm\kappa-i\gamma\ \ (\gamma>0,\
\kappa<\gamma)$. Such states can still have observable
effects, like in the $p+p$ system, because these conjugate
poles are roughly at the same distance from the physical
energies. We have searched for such states in $^3{\rm He}+
{^3{\rm He}}$, and found no unambiguous evidence for their 
existence close to the imaginary $k$ axis.

In summary, we do not find any $S$-matrix poles in either 
$^4{\rm He}+p+p$ or $^3{\rm He}+{^3{\rm He}}$ that could 
cause strong observable effects in the $^3{\rm He}({^3{\rm 
He}},2p){^4{\rm He}}$ cross section. For $^4{\rm He}+p+p$ 
our result is probably the best, one can currently achieve.
Further studies of $^3{\rm He}+{^3{\rm He}}$ in a
$(d+p)+(d+p)$ model or in a full six-body dynamical model
would be interesting.

\section{The $\bbox{^3}$H\lowercase{e}($
\bbox{^3}$H\lowercase{e,2p})$\bbox{^4}$H\lowercase{e}
and $\bbox{^3}$H($\bbox{^3}$H,\lowercase{2n})$
\bbox{^4}$H\lowercase{e} reaction cross sections}

$^3{\rm He}({^3{\rm He}},2p){^4{\rm He}}$ is the only solar
nuclear reaction whose cross section has been measured 
within the  solar Gamow window (around $\approx$$\,20$ keV). 
At such low energies the cross section, measured in the 
laboratory, is enhanced due to screening effects by the 
electrons present in the target atoms \cite{screen}. 
This electron screening effect has definitely been 
identified in the latest LUNA data for $^3{\rm He}({^3{\rm 
He}},2p){^4{\rm He}}$. For applications in the solar models 
the electron screening enhancement has to be subtracted 
from the data.

Currently the $^3{\rm He}(d,p){^4{\rm He}}$ reaction is the 
one for which the enhancement of the low-energy fusion cross 
section due to electron screening is studied best. In 
agreement between experiment \cite{Prati} and theory 
\cite{Shoppa,Langanke,Bang} it appears that the screening 
enhancement for this reaction, in which deuterons collide 
with an atomic $^3$He gas target, is well described in the 
adiabatic limit. In this case the electron screening can be 
represented by a constant shift of the collision energy by 
the screening energy $U_e$ which is given by the difference 
of the electronic binding energy of the united atom and the 
sum of the asymptotic fragments. Applied to the $^3{\rm 
He}+{^3{\rm He}}$ reaction, the screening energy in the
adiabatic limit is $U_e=240$ eV. We will use this value in 
the following, but we note that  fits to the LUNA data
might indicate a somewhat larger screening potential 
\cite{LUNA,Angulo}. These fits had to make  assumptions about the 
energy dependence of the bare-nuclear $^3{\rm He}({^3{\rm 
He}},2p){^4{\rm He}}$ $S$ factor. This has motivated us to 
perform calculations for the bare reaction cross sections 
of $^3{\rm He}({^3{\rm He}},2p){^4{\rm He}}$ and the mirror 
reaction $^3{\rm H}({^3{\rm H}},2n){^4{\rm He}}$.

As we mentioned, a microscopic description of $^3{\rm 
He}({^3{\rm He}},2p){^4{\rm He}}$, which handles the full
three-body final state rigorously, is currently not
feasible. Here we use a simplified version of the
continuum-discretized coupled channel method \cite{cdcc} to
describe $^4{\rm He}+N+N$. In this method the total energy
available for $^4{\rm He}+N+N$ is divided between the
$^4{\rm He}-N$ and $({^4{\rm He}},N)-N$ relative motions in
the $({^4{\rm He}},N)N$ configuration, and between the
$N-N$ and $(N,N)-{^4{\rm He}}$ motions in the $(N,N){^4{\rm
He}}$ configuration. Within the two-cluster subsystems,
$({^4{\rm He}},N)$ or $(N,N)$, the continuum energy is
discretized, and the remaining energy appears as the
scattering energy in the $({^4{\rm He}},N)-N$ and 
$(N,N)-{^4{\rm He}}$ two-body systems. The system of 
coupled channels is built up from the $({^4{\rm He}},N)-N$ 
and $(N,N)-{^4{\rm He}}$ channels containing the various
discretized-energy states in the two-cluster subsystems.

Generally, the discretization of the continuum in the
two-body subsystems is done in equidistant $k$ bins, and
proper continuum states are used. Here we adopt a simpler
approach. We discretize the continuum on a finite
square-integrable basis. Thus, our discretized states are
pseudo-bound states with 
square-integrable wave functions and positive energy. The discretization is
performed by choosing $N_d$ basis states with ranges that
increase following a geometric progression. By varying the
total range of the basis and $N_d$ one can achieve very
different discretization patterns, e.g., dense or sparse
discretization, discretizations including only low energies
or allowing high off-the-energy-shell states also, etc.
This way one can test the sensitivity of the calculated
reaction cross section on the continuum discretization, and
see if this approximation is reasonable or not. We
typically use $N_d=5-10$ and solve the coupled-channel
scattering problem by using the method of Ref.\
\cite{Kamimura}. 

Figure 2 shows our results for $^3{\rm He}({^3{\rm 
He}},2p){^4{\rm He}}$ and $^3{\rm H}({^3{\rm H}},2n){^4{\rm 
He}}$ together with the available experimental data. In
order to get rid of the the trivial exponential dropping \
of the cross sections caused by the Coulomb penetration, 
we use the astrophysical $S$-factor parametrization 
\begin{equation}
S(E)=\sigma (E)E\exp{\Big [2\pi\eta (E)\Big ]}, \hskip 0.5cm
\eta (E)={{\mu Z_1Z_2e^2}\over{k\hbar^2}}.
\end{equation}
Our curves in Fig.\ 2 come from a continuum discretization
that proved to be the most stable at the $({^4{\rm 
He}},N)-N$ and $(N,N)-{^4{\rm He}}$ two-body scattering
level. We also tried other discretization patterns and
found that the absolute normalization of the cross section
curves depend somewhat (10--20\%) on the chosen
discretization, but the shapes of the curves remain very
similar. Our results are close to those of Ref.\
\cite{Desc}, where a similar model was used. Our full model
space is roughly 5--10 times bigger than in Ref.\ 
\cite{Desc}, which allows us to use much more flexible 
continuum discretizations. Nevertheless, all our results 
seem to be similar to Ref.\ \cite{Desc}, e.g., we also 
find that the channels involving the $^4{\rm He}+N$ states 
with $J^\pi=1/2^-$ have a dominant role. 

Fig.\ 3 shows the effect of electron screening on our
calculated $^3{\rm He}({^3{\rm He}},2p){^4{\rm He}}$ $S$
factor with $U_e=240$ eV screening potential. A nice
agreement is observed with the low-energy LUNA data. 

The overall agreement between our results and the 
experimental data is considered to be good. We see,
however, a marked disagreement with the $^3{\rm H}({^3{\rm 
H}},2n){^4{\rm He}}$ data at very low energies. The energy
dependence of our calculated $S$ factor is different from
the precise Los Alamos data. One possible explanation
of this discrepancy could be our approximate treatment of
the three-body final state. However, this is not supported
by our finding that the shape of the $S(E)$ curve is rather
insensitive to the way the discretization is done.
Nevertheless, an improved model with the full three-body
treatment of the final state would be desirable. As we 
mentioned, the thresholds of the ${^4{\rm He}}+n+n$ and 
$^3{\rm H}+{^3{\rm H}}$ channels are too far from each other in 
our model. In order to see if this may affect the 
energy-dependence of the $S$ factor, we made some test 
calculations. We artificially modified the energies of the 
$^3$H clusters to reproduce the experimental threshold 
energy difference. This changed the absolute normalization of 
the $S$ factor (as the $3/2^-$ and especially the $1/2^-$ 
discretized states moved closer to the $^3{\rm H}+{^3{\rm H}}$ 
threshold) but not its shape.

\section{Conclusion}

We have studied the $^3{\rm He}({^3{\rm He}},2p){^4{\rm 
He}}$ and $^3{\rm H}({^3{\rm H}},2n){^4{\rm He}}$ reactions
within the microscopic cluster model using significantly larger model
spaces than previously employed.
Our motivation and results have been twofold:

We searched for signs of possible low-energy resonances in 
$^3{\rm He}({^3{\rm He}},2p){^4{\rm He}}$, which could have
important effects on the $^7$Be and $^8$B solar neutrino
fluxes. The $^3{\rm He}+{^3{\rm He}}$ and $^4{\rm He}+p+p$ 
channels were studied separately, which allowed us to use
methods that can treat the two- and three-body asymptotics
in a rigorous way. We extended the two- and three-body
scattering matrices to complex energies and searched for
their poles. We have not found any indication for the
existence of a low-energy resonance or virtual state that
could cause  observable effects in the cross section.
Thus, it is unlikely that a yet unobserved
resonance around the threshold energy in the
$^3$He($^3$He,2p)$^4$He reaction might affect this important solar
reaction cross section.

We calculated the cross sections of the $^3{\rm He}({^3{\rm 
He}},2p){^4{\rm He}}$ and $^3{\rm H}({^3{\rm H}},2n){^4{\rm 
He}}$ reactions in the continuum-discretized coupled channel
approximation. Our results are in a good general agreement
with available data, except for the very low-energy $^3{\rm 
H}({^3{\rm H}},2n){^4{\rm He}}$ cross section, where we
observe a systematic deviation from the most precise
measurement. Our test calculations show that the
energy dependence of the cross sections is hardly influenced
by the details of the continuum discretization, but might be caused by
the approximate treatment of the 3-body continuum. Here improvements are
certainly warranted.

\acknowledgments

The work of A.\ C.\ was performed under the auspices of the
U.S.\ Department of Energy and was supported by OTKA 
Grant F019701 and by the Bolyai Fellowship of the Hungarian 
Academy of Sciences. We also thank the Danish Research Council 
and the Theoretical Astrophysics Center for financial support.

\begin{figure}
\epsfxsize 16cm \epsfbox{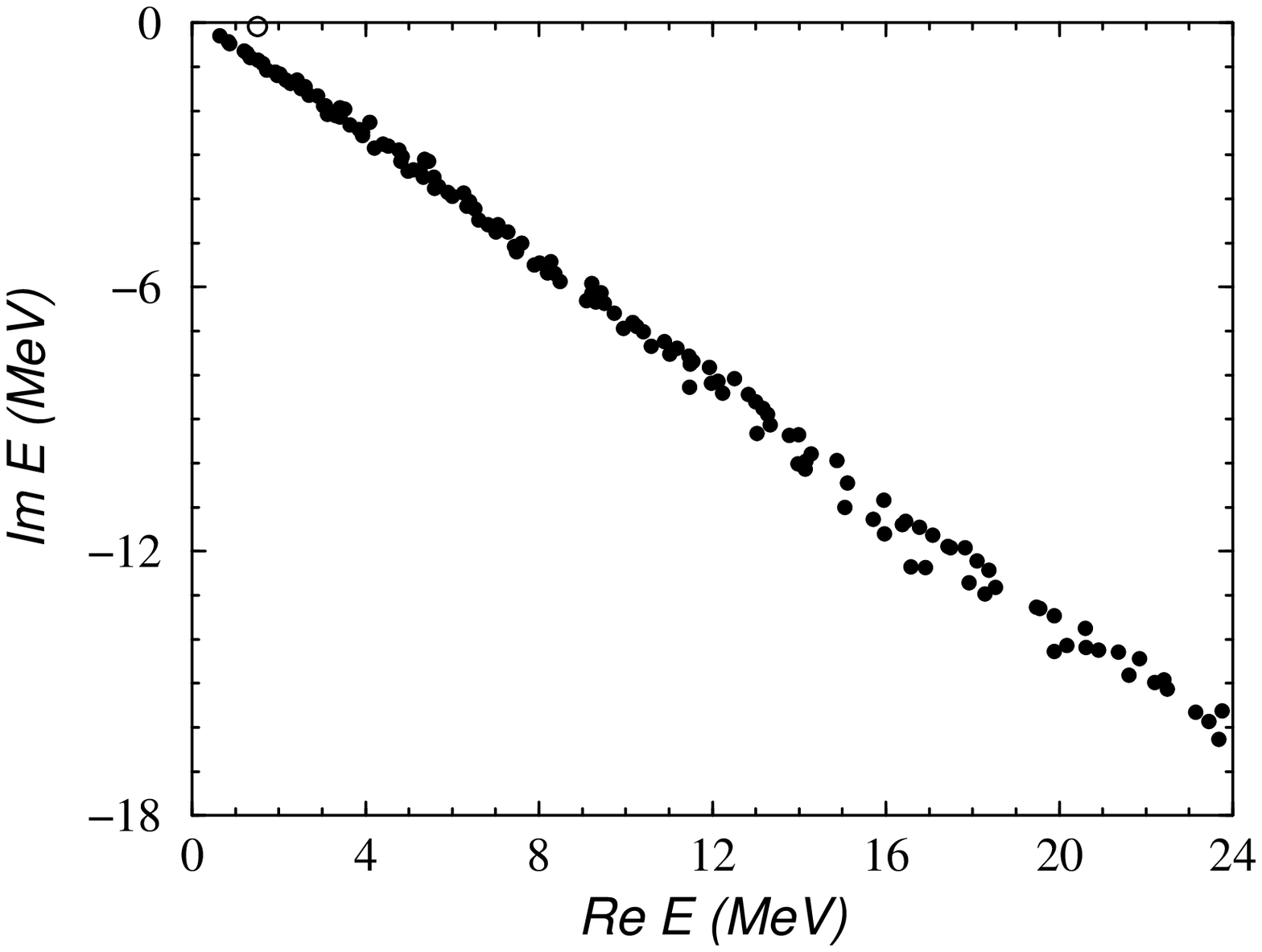}
\vskip 1cm
\caption{Energy-eigenvalues of the complex scaled
Hamiltonian of the $J^\pi=0^+$ states in $^4{\rm He}+p+p$. 
The dots are the points of the rotated discretized continuum,
while the circle is the ground state resonance of $^6$Be.}
\label{Fig1}
\end{figure}
\newpage

\begin{figure}
\epsfxsize 16cm \epsfbox{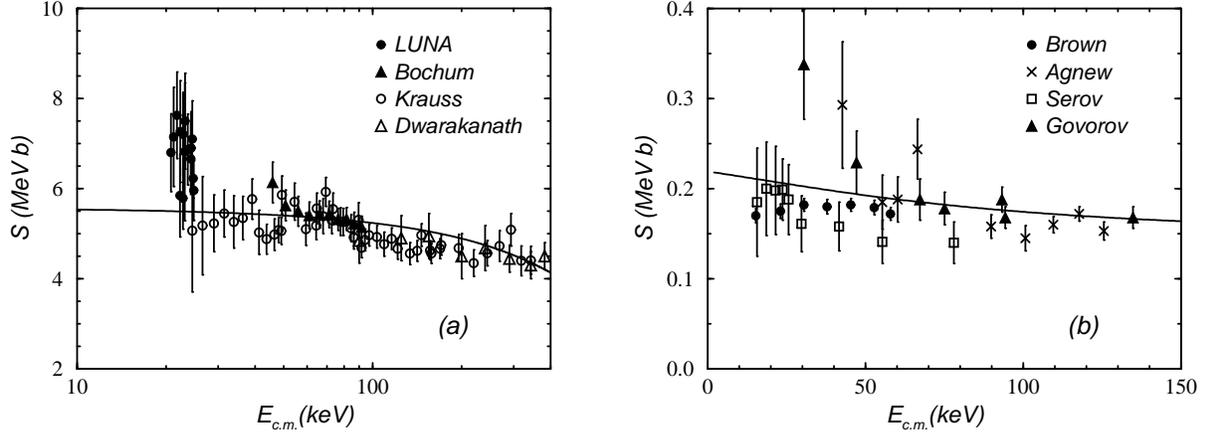}
\vskip 1cm
\caption{Astrophysical $S$ factors for the (a) $^3{\rm 
He}({^3{\rm He}},2p){^4{\rm He}}$ and (b) $^3{\rm 
H}({^3{\rm H}},2n){^4{\rm He}}$ reactions. The
experimental data are taken from (a) \protect\cite{LUNA}
(filled circle, and filled triangle), \protect\cite{Krauss}
(open circle), \protect\cite{Dw} (open triangle) and (b) 
\protect\cite{Brown} (filled circle), \protect\cite{Agnew}
(cross), \protect\cite{Serov} (square), and
\protect\cite{Govorov} (triangle). The solid curves are our
results.}
\label{Fig2}
\end{figure}
\newpage

\begin{figure}
\epsfxsize 16cm \epsfbox{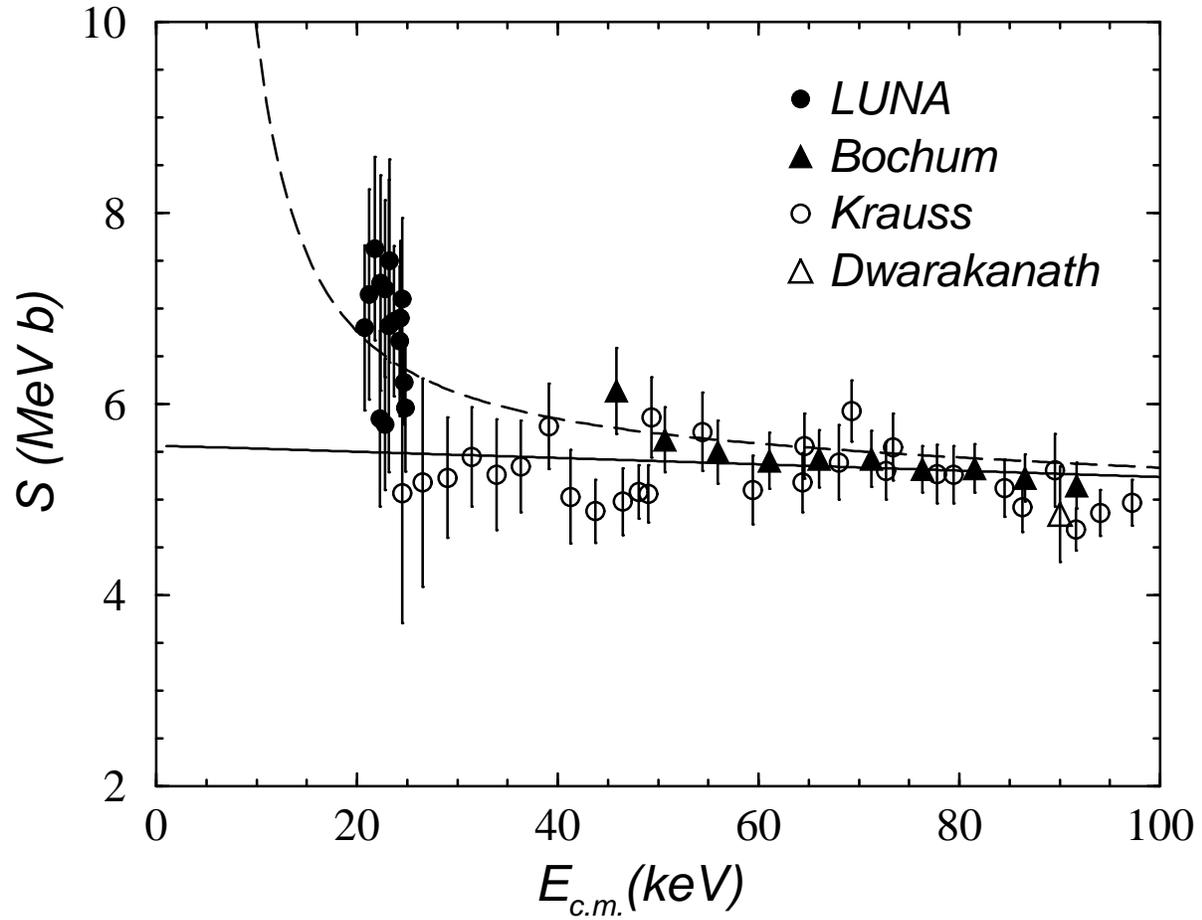}
\vskip 1cm
\caption{The same as Fig.\ 2(a), except that the effect of 
electron screening on our low-energy theoretical curve is 
shown by the dashed line. The adiabatic screening potential, 
$U_e=240$ eV, is used.}
\label{Fig3}
\end{figure}

\end{document}